# Non static Global monopole in Lyra geometry

F. Rahaman and R. Mondal

Department of Mathematics

Jadavpur University, Kolkata – 700 032, India

E-mail: farook_rahaman@yahoo.com

Abstract:
A class of non static solutions around a global monopole resulting from the breaking of a global S0(3) symmetry based on Lyra geometry are obtained. The solutions are obtained using the functional separability of the metric coefficients. We have shown that the monopole exerts attractive gravitational effects on test particles.

## Introduction:

Global monopoles, predicted to exist in Grand Unified Theory, are supposed to have been created during phase transition in the early Universe [1]. They are stable topological defects produced when global S0(3) symmetry is spontaneously broken in U(1). Monopoles exhibit some interesting properties, particularly in relation to the appearance of non trivial space time topologies [1, 2]. Using a suitable scalar field it can be shown that spontaneous symmetry breaking can give rise to such objects which are nothing but the topological knots in the vacuum expectation value of the scalar field and most of their energy is concentrated in a small region near the monopole core. From the topological point of view they are formed in the vacuum manifold M when M contains surfaces which can not be continuously shrunk to a point i.e. when $\pi_2(M) \neq I$. Such monopoles have Goldstone fields with energy density decreasing with the distance as inverse square law. They are also found to have some interesting features in the sense that a monopole exerts no gravitational force on its surrounding non relativistic matter but space around it has a deficit solid angle [2].

At first , Barriola and Vilenkin (BV)[3] showed the existence of such a monopole solution resulting from the breaking of global S0(3) symmetry of a triplet scalar field in a Schwarzchild back ground. After that so many works have been done on general relativistic static models of the global monopole space time [4]. In recent past, Chakraborty [5,6] and Farook [7] have derived the solutions to the Einstein's field equations for the non static space time metric outside the core of a global monopole.

In last few decades there has been considerable interest in alternative theory of gravitation. The most important among them being scalar tensor theories proposed by Lyra [8] and Brans-Dicke [8]. Lyra proposed a modification of Riemannian geometry by introducing a gauge function into the structure less manifold that bears a close resemblances to Weyl's geometry. In general relativity, Einstein succeeded in geometrising gravitation by identifying the metric tensor with gravitational potentials. In scalar tensor theory of Brans-Dicke on the other hand, the scalar field remains alien to the geometry. Lyra's geometry is more in keeping with the spirit of Einstein's principle of geometrisation since both the scalar and tensor fields have more or less intrinsic geometrical significance.

In consecutive investigations Sen [9] and Sen and Dunn [9] proposed a new scalar tensor theory of gravitation and constructed an analog of the Einstein field equation based on Lyra's geometry which in normal gauge may be written as

$$R_{ik} - \tfrac{1}{2} g_{ik} R + (3/2) \phi_i \phi_k - \tfrac{3}{4} g_{ik} \phi_m \phi^m = -8\pi G T_{ik} \qquad \ldots(1)$$

where $\phi_i$ is the displacement vector and other symbols have their usual meaning as in Riemannian geometry.

Halford [10] has pointed out that the constant displacement field $\phi_i$ in Lyra's geometry play the role of cosmological constant $\Lambda$ in the normal general relativistic treatment. According to Halford, the present theory predicts the same effects within observational limits, as far as the classical solar system tests are concerned, as well as tests based on the linearized form of field equations. Soleng [11] has pointed out that the constant displacement field in Lyra's geometry will either include a creation field and be equal to Hoyle's creation field cosmology or contain a special vacuum field which together with the gauge vector term may be considered as a cosmological term.

Subsequent investigations were done by several authors in scalar tensor theory and cosmology within the frame work of Lyra geometry [12]. Recently, Rahaman et al have studied some topological defects within the frame work of Lyra geometry[13].

In this work, we shall deal with monopole with constant displacement vectors based on Lyra geometry in normal gauge i.e. displacement vector

$$\phi_i = (\beta, 0, 0, 0) \qquad \ldots\ldots(2)$$

and look forward whether the monopole shows any significant properties due to introduction of the gauge field in the Riemannian geometry.

## 2. The basic equations:

Here we closely follow the formalism of Chakraborty [6] and take the Lagrangian that gives rise to monopoles as

$$L = \tfrac{1}{2} g^{\mu\gamma} \partial_\mu \Phi^a \partial^\gamma \Phi^a - \tfrac{1}{4} \lambda (\Phi^a \Phi^a - \eta^2)^2 \qquad \ldots(3)$$

where $\Phi^a$ is the triplet scalar field $a = 1,2,3$ and $\eta$ is the energy scale of symmetry breaking. For non static monopole, we do not write the explicit form of the field configuration of $\Phi^a$ but take it as implicit form.

The energy momentum tensor for the above Lagrangian is given by [6]

$$T_\mu^{\ \gamma} = \nabla_\mu \Phi^a \cdot \nabla^\gamma \Phi^a - L \delta_\mu^{\ \gamma} \qquad \ldots\ldots(4)$$

The metric ansatz describing a monopole can be written as

$$ds^2 = -A\,dt^2 + B\,dr^2 + C\,d\Omega_2^2 \qquad \ldots(5)$$

Here, A, B and C are functions of r and t.

The field equations (1) for the metric (5) reduces to

$$(\tfrac{1}{2}B)[\,2(C^{11}/C) + \tfrac{1}{2}(C^1/C)^2 - (B^1 C^1/BC)\,] - (\tfrac{1}{2}A)[\,\tfrac{1}{2}(C^{\bullet 2}/C)^2 + (B^\bullet C^\bullet/BC)\,]$$

$$- (1/C) + \tfrac{3}{4}(1/A)\beta^2 = \tfrac{1}{2}[-(1/A)(\Phi^{a\bullet})^2 - (1/B)(\Phi^{a1})^2 + \tfrac{1}{2}\lambda(\Phi^a \Phi^a - \eta^2)^2] \qquad \ldots(6)$$

$$(\tfrac{1}{2}B)[\,(C^1/C)^2 + (A^1 C^1/AC)\,] - (\tfrac{1}{2}A)[\,2(C^{\bullet\bullet}/C) + \tfrac{1}{2}(C^\bullet/C)^2 - (A^\bullet C^\bullet/AC)\,]$$

$$- (1/C) - \tfrac{3}{4}(1/A)\beta^2 = \tfrac{1}{2}[\,(1/A)(\Phi^{a\bullet})^2 + (1/B)(\Phi^{a1})^2 + \tfrac{1}{2}\lambda(\Phi^a \Phi^a - \eta^2)^2\,] \qquad ..(7)$$

$$(\tfrac{1}{2}B)[\,(C^{11}/C) + (A^{11}/A) + \tfrac{1}{2}(A^1/A)^2 + \tfrac{1}{2}(C^1/C)^2 - \tfrac{1}{2}(B^1 C^1/BC) -$$

$$(B^1 A^1/BA) + A^1 C^1/2AC\,]$$

$$+ (\tfrac{1}{2}A)[\,-(C^{\bullet\bullet}/C) - (B^{\bullet\bullet}/B) - \tfrac{1}{2}(C^{\bullet 2}/C)^2 - \tfrac{1}{2}(B^{\bullet 2}/B)^2 + \tfrac{1}{2}(A^\bullet C^\bullet/AC)$$

$$+ \tfrac{1}{2}(A^\bullet B^\bullet/AB) - \tfrac{1}{2}(B^\bullet C^\bullet/BC)\,] - \tfrac{3}{4}(1/A)\beta^2$$

$$= \tfrac{1}{2}[\,(1/A)(\Phi^{a\bullet})^2 - (1/B)(\Phi^{a1})^2 + \tfrac{1}{2}\lambda(\Phi^a \Phi^a - \eta^2)^2\,] \qquad \ldots(8)$$

$$-(C^{\bullet 1}/C) - \tfrac{1}{2}(C^\bullet C^1/C^2) + \tfrac{1}{2}(A^1 C^\bullet/AC) + \tfrac{1}{2}(B^\bullet C^1/BC) = \Phi^{a\bullet}.\Phi^{a1} \qquad \ldots\ldots(9)$$

The field equation for the scalar triplet $\Phi^a$ is

$$(1/A)[\,-(\Phi^{a\bullet\bullet}/\Phi^a) + (\Phi^{a\bullet}/\Phi^a)\{-(A^\bullet/A) + (B^\bullet/B) + 2(C^\bullet/C)\}\,] +$$

$$(1/B)[\,-(\Phi^{a11}/\Phi^a) + (\Phi^{a1}/\Phi^a)\{-(A^1/A) + (B^1/B) - 2(C^1/C)\}\,]$$

$$+ \lambda(\Phi^a \Phi^a - \eta^2) = 0 \qquad \ldots\ldots(10)$$

[ The symbols '$^\bullet$' and '$^1$' represent the differentiation with respect to 't' and 'r' respectively ]

## 3. Solutions to the field equations:

As the field equations are complicated to solve the field equations, we shall assume the separable form of the metric coefficients as follows:

$$A = A_1(r).A_2(t) ; B = B_1(r).B_2(t) ; C = C_1(r).C_2(t) \qquad ...(11)$$

Further, without any loss of generality, one can assume

$$A_2(t) = B_1(r) = 1 \qquad ....(12)$$

[$A_2(t)$ or $B_1(r)$ different from unity results in a scaling of time or radial coordinates]

Also we have taken the scalar field triplet in the separable form as

$$\Phi^a(r,t) = \Phi^a_1(r) + \Phi^a_2(t) \qquad ........(13)$$

We shall now solve these equations with the following relations among the metric coefficients:

$$A_1 = aC_1^n \text{ and } B_2 = bC_2^m \qquad .........(14)$$

[ where a,b,m,n are arbitrary constants]

From eq.(9) by using eqs.(11) – (14), we get

$$\Phi^{a\bullet}_2 = q(m + n - 3/2)(C^\bullet_2/C_2) \text{ and } \Phi^{a\prime}_1 = (1/q)(C_1'/C_1) \qquad ......(15)$$

where q is the separation constant.

Now eliminating $\Phi^{a\prime}_1$ and using the separable forms and following the relations

(6) + (7) – 2 (8), we get

$$(C_1''/C_1) + d(C_1'/C_1)^2 = e\, C_1^{-n} \qquad .......(16)$$

where $d = \frac{1}{2}[3n^2 - n + (2/q^2)]$ ; $e = (p/2a)$ and p is separation constant.

The integral form of $C_1$ is

$$\int [D_1 C_1^{-2d} + \{2e/(2d - n + 2)\}C_1^{(2-n)}]^{-\frac{1}{2}} dC_1 = \pm(r - r_0) \qquad .......(17)$$

where $D_1$ and $r_0$ are integration constants.

For a different choice of the constants the solutions for $C_1$ are

Case-I : $p = 0$ : $C_1 \propto (r - r_0)^{[1/(1+d)]}$ ......(18)

Case-II : $D_1 = 0$ : $C_1 \propto (r - r_0)^{[2/n]}$ ......(19)

Case-III : $n = 0$, $D_1 > 0$ :

$C_1 = \sqrt{[(d+1)(D_1/e)]} [\sinh \sqrt{\{e(d+1)\}}(r - r_0)]^{[1/(d+1)]}$ .....(20)

Case-IV : $n = 0$, $D_1 < 0$ :

$C_1 = \sqrt{[(d+1)(D_1/e)]} [\cosh \sqrt{\{e(d+1)\}}(r - r_0)]^{[1/(d+1)]}$ .....(21)

Proceeding in a similar way, the integral form of $C_2$ is

$\int [D_2 C_2^{-2f} + \{2g/(2f-m+2)\}C_2^{(2-m)} - \{6\beta^2/(2m+4)(2f+2)\}C_2^2]^{-\frac{1}{2}} dC_2 = \pm (t - t_0)$ ..(22)

[ where $D_2$ and $t_0$ are integration constants and $g = p/b(2m+4)$,

$f = [\{3m^2 + 2 + 2q^2(m+n-3/2)^2\}/(2m+4)]$ ]

The solution set for the time part $C_2$ is as follows:

Case-I : $p = 0$ :

$C_2 = \sqrt{[\beta^2 \{(f+1)/(2m+4)\}]} [\sin\sqrt{\{2 D_2 (f+1)(2m+4)/\beta^2\}}(t - t_0)]^{[1/(f+1)]}$ ...(23)

Case-II : $D_2 = 0$, $m = 0$ :

$C_2 = \exp[\sqrt{F}(t - t_0)]$ where $F = [(g/2f) - 3\{\beta^2/8(f+1)\}]$ ........(24)

Case-III : $D_2 = 0$, $m = 2$ :

$C_2 = \sqrt{(2g/3\beta^2)} [\sin \sqrt{\{3\beta^2/(4(f+1))\}}(t - t_0)]$ ......(25)

Case-IV : $f = 1$, $m = 0$ :

$C_2^2 = 2\sqrt{[D_2/\{g - (3/2)\beta^2\}]} [\sinh \sqrt{\{g - (3/2)\beta^2\}}(t - t_0)]$ ......(26)

From eq.(15), we get

$\Phi^a_1(r) = (1/q)\ln C_1 + \Phi^a_{01}$ and $\Phi^a_2(t) = q[(m+n-(3/2))]\ln C_2 + \Phi^a_{02}$ ....(27)

(with $\Phi^a_{01}$ and $\Phi^a_{02}$ are integration constants.)

## 4. Concluding Remarks:

In this paper, we have studied the gravitational field of non static monopole in Lyra geometry. The solutions we have obtained in the present paper are not most general. But nevertheless the solutions presented here perhaps the exact analytical solutions obtained for the first time.

The expression of our metric (5) is

$$ds^2 = -C_1^n \, dt^2 + C_2^m \, dr^2 + C_1 C_2 \, d\Omega_2^2$$

If we define $T = \int (C_2)^{-(m/2)} dt$ and $R = \int (C_1)^{(-n/2)} dr$

Then the above metric can be written as

$$ds^2 = -dT^2 + dR^2 + C_1^{(1-n)} C_2^{(1-m)} d\Omega_2^2$$

The metric describes a solid angle of deficiency, which depends both on radial and time Coordinates ( except for a conformal factor ).
It is important to note that our non static metric is not conformally flat and hence it represents a monopole [6]

Another aspect of the monopole is the effect on test particle in its gravitational field. Let us consider an observer with four velocity given by

$$V_i = \sqrt{C_1} \, \delta_i^t .$$

Then we obtain the acceleration vector $A^i$ as

$$A^i = V^i_{;k} V^k = (C_1^1 / C_1) C_1^{-2} \delta_r^i .$$

For the above solutions (18) – (21), one can see that $A^r$ is positive. Hence the monopole exhibits an attractive nature to the observer .

## Acknowledgements:

We are thankful to the members of Relativity and Cosmology Center, Jadavpur University for helpful discussions. F.R is thankful to DST, Govt. of India for financial support.